\documentclass[a4paper]{article}
\usepackage{ASVspoof2021}
\usepackage{epsfig,amssymb,amsmath}

\usepackage{multirow}
\usepackage{amsmath}
\usepackage{color}
\usepackage{amssymb, bm, mathtools,array}
\usepackage{dsfont}
\usepackage{verbatim}

\usepackage{url}

\newcommand{\tabincell}[2]{\begin{tabular}{@{}#1@{}}#2\end{tabular}}

\ninept

\title{Multi-Task Learning in Utterance-Level and Segmental-Level Spoof Detection}

\makeatletter
\def\name#1{\gdef\@name{#1\\}}
\makeatother
\name{{\em Lin Zhang$^{1,2}$, Xin Wang$^1$, Erica Cooper$^1$, Junichi Yamagishi$^{1,2}$}}
\address{
  $^1$National Institute of Informatics, Tokyo, Japan \\
  $^2$SOKENDAI (The Graduate University for Advanced Studies), Kanagawa, Japan\\
{\small \tt \{zhanglin, wangxin, ecooper, jyamagis\}@nii.ac.jp }}

\begin{document}
\maketitle

\begin{abstract}

In this paper, we provide a series of multi-tasking benchmarks for simultaneously detecting spoofing at the segmental and utterance levels in the PartialSpoof database. First, we propose the SELCNN network, which inserts squeeze-and-excitation (SE) blocks into a light convolutional neural network (LCNN) to enhance the capacity of hidden feature selection. Then, we implement multi-task learning (MTL) frameworks with SELCNN followed by bidirectional long short-term memory (Bi-LSTM) as the basic model. We discuss MTL in PartialSpoof in terms of architecture (uni-branch/multi-branch) and training strategies (from-scratch/warm-up) step-by-step. Experiments show that the multi-task model performs relatively better than single-task models. Also, in MTL, a binary-branch architecture more adequately utilizes information from two levels than a uni-branch model. For the binary-branch architecture, fine-tuning a warm-up model works better than training from scratch. Models can handle both segment-level and utterance-level predictions simultaneously overall under a binary-branch multi-task architecture. Furthermore, the multi-task model trained by fine-tuning a segmental warm-up model performs relatively better at both levels except on the evaluation set for segmental detection. Segmental detection should be explored further.



\end{abstract}

\section{Introduction}\label{sec:intro}
Automatic speaker verification (ASV) \cite{kinnunen2010overview} systems are for verifying the identity of individual speakers. ASV is a convenient and efficient biometric authentication method that has many practical applications. However, ASV systems are vulnerable to circumvention by spoofing attacks, more formally known as “presentation attacks” \cite{jain2006biometrics}. This attack scenario was first discussed in a special session of Interspeech 2013 \cite{evans2013spoofing}. Since then, the ASVspoof challenge \cite{Wu2014, Kinnunen2017, Todisco2019} has been held biennially since 2015 to investigate protecting automatic speaker verification from attacks. Until now, several attacks have been explored in this challenge, including logical access (LA), physical access (PA), and a new Deepfake (DF) scenario \cite{asvspoof2021} introduced in 2021. The LA attacks are generated using the latest voice conversion (VC) and text-to-speech (TTS) techniques. The PA attacks are created using replayed samples in simulated or real environments, and the DF task involves compressed synthesized audio. However, all of those attacks have overlooked another realistic scenario: {\em partially-spoofed} audio in which spoofed utterances can contain bona fide segments that will degrade countermeasure performance \cite{Zhang2021}.

To explore this realistic scenario in depth, we have designed the PartialSpoof \cite{Zhang2021} database, based on the LA portion of ASVspoof 2019 \cite{Wang2020data}. In PartialSpoof, utterances can consist of a combination of spoofed and bona fide segments. Injected spoof slices are as short as 10 ms, and spoof segment ratios\footnote{Spoof segment ratio is the ratio of the total duration of spoofed segments within an entire length of audio.} range from 0.23\% to 99.81\%. We constructed separate utterance-level and segmental-level models for the database, and preliminarily discussed detection at the two levels. Results show that detecting spoofed segments injected into an utterance is an understandably challenging task. \cite{Yi2021halftruth} stated the same regarding a similar database generated by the LPCNet \cite{valin2018lpcnet} model and a Mandarin speech corpus (AISHELL-3 \cite{shi2020aishell3}) in the same period. 

Furthermore, it is apparent that information at the segmental level and the utterance level are related because segments are components of an utterance, and utterance labels can be uniquely determined by a sequence of segmental labels. We realize that simply exploring independent models for segmental- and utterance-level spoofing detection (as in our previous work \cite{Zhang2021}) is inadequate. Some questions remain: \textit{Can we train one model that can be used for both utterance-level detection and segmental detection? Can those two dependent types of labels each help to improve the accuracy of the other label type?}

Considering the relationship between segments and utterances, multi-task learning (MTL) can be used to answer the above questions. MTL leverages useful information contained in multiple related tasks to help improve generalization of all the tasks \cite{caruana1997multitask,Zhang2021multi}. MTL is effective in speech fields such as anti-spoofing and speaker verfication \cite{Li2019asvspoofMTLaccess}, speaker recognition \cite{wang2019usage}, speech recognition and machine translation \cite{indurthi2021task}, voice conversion and speech synthesis \cite{wu2015deep, CHEN2021101243mtl}, voice activity detection and speech enhancement \cite{tan2021speech}, paralinguistic detection \cite{pan2021multi}, and many others.

In contrast to the aforementioned multi-task problems, segments and utterances have a more direct relationship. Hence, we discuss MTL for PartialSpoof in two parts: structure (uni-branch vs.\ binary-branch) and training strategies (from-scratch vs.\ warm-up). Results indicate that an appropriate multi-task method performs efficient multi-level spoof detection. The binary-branch structure aided by using the segmental model for warm-up (SegBW) performed relatively better than the other models. Overall, this study provides a series of benchmarks for simultaneous detection at the segmental and utterance levels for the PartialSpoof database.

This paper is organized as follows: In Section \ref{sec:CM}, we present the upgraded basic model SELCNN used throughout the rest of the paper. In Section \ref{sec:single_multi}, we compare single-task and multi-task learning. In Section \ref{sec:train_multi}, we further explore training strategies for MTL. In Section \ref{sec:exp}, we detail the experiments. In Section \ref{sec:res}, we discuss the results. In Section \ref{sec:conclusion}, we conclude this paper.

\section{Basic Model}\label{sec:CM}

\begin{figure}[t]
  \centering
    \includegraphics[width=0.66\linewidth]{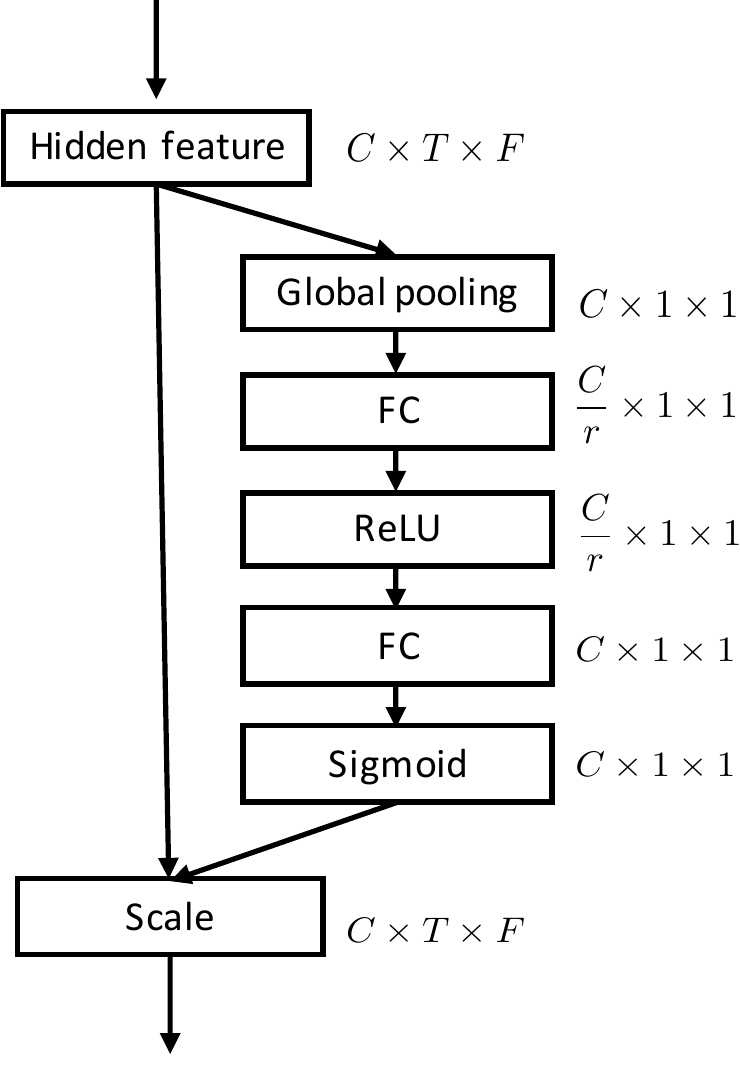}
  \caption{\it Architecture of SE block. $C$, $T$, $F$ represent dimension of Channel, Time and Frequency respectively. $r$ represents reduction ratio.}
  \label{fig:se}
\end{figure}

We investigated a variant of light convolutional neural networks (LCNN) for PartialSpoof\footnote{Table 1 of \cite{Zhang2021}: LCNN + AP, LCNN + SAP, LCNN + Bi-LSTM + AP, LCNN + Bi-LSTM + SAP} in our previous work \cite{Zhang2021}. LCNNs \cite{wu2018light, Lavrentyeva2019} are widely used in anti-spoofing because they are light and robust. The key component in an LCNN is the Max-Feature-Map (MFM) operation based on the Max-Out activation function; it plays a local feature selection role by suppressing the activation of some neurons in the channel axis. It can be seen as selecting the optimal feature at each location learned by different filters. However, the Max-Out activation function is too coarse-grained for channel selection; thus, we introduce squeeze-and-excitation (SE) blocks \cite{Hu2020se} into the LCNN to aid in learning the attention weight for channels. This is what we call `SELCNN.' As Table \ref{tab:selcnn} shows, we insert an SE block (in red) before 
each convolution layer (except Conv\_0)\footnote{We also tried other positions: \textit{between Conv and MFM} yields a worse performance; \textit{after MFM} shows a similar performance with \textit{before Conv}.}. The SE block structure is shown in Figure \ref{fig:se}.

\begin{table}[th]
\caption{\it Structure of SELCNN. [B, C, T, F] are [Batch size, Channel, Time, Frequency] respectively.)}
\label{tab:selcnn}
\setlength{\tabcolsep}{0.9mm}{
\centerline{
\begin{tabular}{|c|c|c|}
\hline
\textbf{Type} & \textbf{Filter/Stride/Padding} & \multicolumn{1}{c|}{\textbf{Output  Size{[}B, C, T, F{]} }} \\
\hline \hline
Conv\_0       & 5 x 5 / 1 x 1 / 2              & {[}B, 64, T, F{]}                                                                                                          \\
MFM\_1        & -                              & {[}B, 32, T, F{]}                                                                                                          \\
MaxPool\_2    & 2 x 2 / 2 x 2 / 0              & {[}B, 32, T // 2, F // 2{]}                                                                                                \\
\textcolor{red}{SE block}      & -             & {[}B, 32, T // 2, F // 2{]}                                                                                                \\
Conv\_3       & 1 x 1 / 1 x 1 / 0              & {[}B, 64, T // 2, F // 2{]}                                                                                                \\
MFM\_4        & -                              & {[}B, 32, T // 2, F // 2{]}                                                                                                \\
BatchNorm\_5  & -                              & {[}B, 32, T // 2, F // 2{]}                                                                                                \\
\textcolor{red}{SE block}      & -             & {[}B, 32, T // 2, F // 2{]}                                                                                                \\
Conv\_6       & 3x 3/ 1 x 1 / 1                & {[}B, 96, T // 2, F // 2{]}                                                                                                \\
MFM\_7        & -                              & {[}B, 48, T // 2, F // 2{]}                                                                                                \\
MaxPool\_8    & 2 x 2 / 2 x 2 / 0              & {[}B, 48, T // 4, F // 4{]}                                                                                                \\
BatchNorm\_9  & -                              & {[}B, 48, T // 4, F // 4{]}                                                                                                \\
\textcolor{red}{SE block}      & -             & {[}B, 48, T // 4, F // 4{]}                                                                                                \\
Conv\_10      & 1 x 1 / 1 x 1 / 0              & {[}B, 96, T // 4, F // 4{]}                                                                                                \\
MFM\_11       & -                              & {[}B, 48, T // 4, F // 4{]}                                                                                                \\
BatchNorm\_12 & -                              & {[}B, 48, T // 4, F // 4{]}                                                                                                \\
\textcolor{red}{SE block}      & -             & {[}B, 48, T // 4, F // 4{]}                                                                                                \\
Conv\_13      & 3x 3/ 1 x 1 / 1                & {[}B, 128, T // 4, F // 4{]}                                                                                               \\
MFM\_14       & -                              & {[}B, 64, T // 4, F // 4{]}                                                                                                \\
MaxPool\_15   & 2 x 2 / 2 x 2 / 0              & {[}B, 64, T // 8, F // 8{]}                                                                                                \\
\textcolor{red}{SE block}      & -             & {[}B, 64, T // 8, F // 8{]}                                                                                                \\
Conv\_16      & 1 x 1 / 1 x 1 / 0              & {[}B, 128, T // 8, F // 8{]}                                                                                               \\
MFM\_17       & -                              & {[}B, 64, T // 8, F // 8{]}                                                                                                \\
BatchNorm\_18 & -                              & {[}B, 64, T // 8, F // 8{]}                                                                                                \\
\textcolor{red}{SE block}      & -             & {[}B, 64, T // 8, F // 8{]}                                                                                                \\
Conv\_19      & 3x 3/ 1 x 1 / 1                & {[}B, 64, T // 8, F // 8{]}                                                                                                \\
MFM\_20       & -                              & {[}B, 32, T // 8, F // 8{]}                                                                                                \\
BatchNorm\_21 & -                              & {[}B, 32, T // 8, F // 8{]}                                                                                                \\
\textcolor{red}{SE block}      & -             & {[}B, 32, T // 8, F // 8{]}                                                                                                \\
Conv\_22      & 1 x 1 / 1 x 1 / 0              & {[}B, 64, T // 8, F // 8{]}                                                                                                \\
MFM\_23       & -                              & {[}B, 32, T // 8, F // 8{]}                                                                                                \\
BatchNorm\_24 & -                              & {[}B, 32, T // 8, F // 8{]}                                                                                                \\
\textcolor{red}{SE block}      & -             & {[}B, 32, T // 8, F // 8{]}                                                                                                \\
Conv\_25      & 3x 3/ 1 x 1 / 1                & {[}B, 64, T // 8, F // 8{]}                                                                                                \\
MFM\_26       & -                              & {[}B, 32, T // 8, F // 8{]}                                                                                                \\
MaxPool\_27   & 2 x 2 / 2 x 2 / 0              & {[}B, 32, T // 16, F // 16{]}                                                                                              \\
Dropout\_28   & -                              &                                          \\                                                                    
\hline
\end{tabular}}}
\end{table}

The remaining structures after \textit{Dropout\_28} are the same as \cite{Zhang2021}: We feed the output of SELCNN to two Bi-LSTM layers, and we set the residual structure over two Bi-LSTM layers to stabilize the training process. Then, we use one fully-connected layer (FC), and apply average pooling (AP) before FC in the utterance-based model. \textit{MSE for P2SGrad} \cite{wang2021comparative} is used as the loss function because it is hyper-parameter-free and performs well in anti-spoofing tasks. 


\section{Single-task and multi-task learning}\label{sec:single_multi}
As discussed in Section \ref{sec:intro}, we aim to build one model that can perform both utterance-level detection and segmental-level detection. Although some conventional CMs can produce scores at both levels during inference \cite{Zhang2021}, these CMs are trained using either segment or utterance labels. Such a single-task learning strategy ignores the relationship between segment and utterance labels. Therefore, we propose using multi-task training to further boost the system performance. This section briefly explains both conventional single-task and proposed MTL strategies.

\begin{table*}[ht]
\caption{\label{tab:eq}{\it Loss functions and predicted scores for basic models of different levels. $\mathcal{D}$ is the training data set. $C=2$ is the number of target classes. $k=1$ is \textit{bona fide}, and $k=2$ is \textit{spoofed}. Superscripts denote the basic level of the trained model: $(utt)$ for the utterance-trained model and $(seg)$ for the segmental-trained model. $s_j$ represents the score of the $j$-th trial, and $s_{j,m}$ represents the score of the $m$-th segment in the $j$-th trial. $\mathds{1}(\cdot)$ is an indicator function. $\tilde{w}_m^{(j)} = 1 / M_j$ because we applied AP. }}
\vspace{2mm}
\centerline{
\begin{tabular}{|c|c|c|c|}
\hline
\textbf{Basic}  &  \textbf{Loss}    & \multicolumn{2}{c|}{\textbf{Score}}                 \\
\cline{3-4}
\textbf{Model}  &  \textbf{Function} & \textbf{Utterance-level}  & \textbf{Segmental-level}  \\
\hline  \hline
\textbf{Utterance} 
& $\mathcal{L}^{(\text{utt-p2s})} = \frac{1}{|\mathcal{D}|}\sum_{j=1}^{|\mathcal{D}|}\sum_{k=1}^{C} (\cos\theta_{j,k} - \mathds{1}(y_j=k)) ^ 2$  
&    \tabincell{r}{$s_j^{(utt)} = \cos\theta_{j,1}$ \\  $=\widehat{\boldsymbol{c}}_1^\top\widehat{\boldsymbol{o}}_j$  }                   
& \tabincell{l}{$s_{j,m}^{(utt)} = \tilde{w}_m^{(j)} \cos\theta_{j, 1, m}$ \\
 $=\tilde{w}_m^{(j)}\widehat{\boldsymbol{c}}_1^{\top}\widehat{\boldsymbol{h}}_{m}^{(j)}$\\
 $=\frac{||\boldsymbol{h}_m^{(j)} ||}{||\boldsymbol{o}_j||} \widehat{\boldsymbol{c}}_1^{\top}\widehat{\boldsymbol{h}}_{m}^{(j)}$
 }                       \\
 
\hline
\textbf{Segmental} 
& $\mathcal{L}^{(\text{seg-p2s})} = \frac{1}{|\mathcal{D}|}\frac{1}{M_j}\sum_{j=1}^{|\mathcal{D}|}\sum_{m=1}^{M_j}\sum_{k=1}^{C} (\cos\theta_{j,k,m} - \mathds{1}(y_{j,m}=k)) ^ 2$                             
& $s_j^{(seg)} = \min_{m} {s_{j,m} }$                        
&   $s_{j,m}^{(seg)}=\cos\theta_{j,1,m}$     \\               
\hline  
\end{tabular}}
\end{table*}

\subsection{Single-Task Learning}
We used single-task learning to train utterance-basic and segmental-basic models based on SELCNN + BiLSTM architecture using either utterance- or segment-level labels. The main differences between them are the target labels and the AP layer. The pooling layer of the utterance-basic model is used to convert the segmental feature matrix into an utterance-level vector, which can also handle varied-length input to the neural network. In comparison, the segmental-basic model omits the pooling layer to keep all observation vectors for each segment and is trained using the segmental labels. Models in our previous study \cite{Zhang2021} are grouped in this category. So, models that use single-task learning are trained completely independently with either utterance- or segmental-level labels.

Corresponding loss functions and equations for predicted scores are described in Table \ref{tab:eq}. In this table, we use $\boldsymbol{x}_{1:N^{(j)}} \in \mathbb{R}^{N^{(j)}\times{D}}$ to represent the input for the $j$-th utterance with $N^{(j)}$ frames, and the input feature of each frame has $D$ dimensions. Then, we define utterance-level labels as $y_j$ and segmental-level labels as $y_{j,m}$, where $m\in[1, N^{(j)}]$ is the frame index. $\cos \theta_{j,k}$ refers to the cosine similarity between the length-normed hidden features for the $j$-th trial and the $k$-th class vector ($k=1$ refers to bona fide, and $k=2$ is for spoof). Each separate model is detailed below.

\begin{figure*}[thb]
 \centering
 \includegraphics[width=1\linewidth]{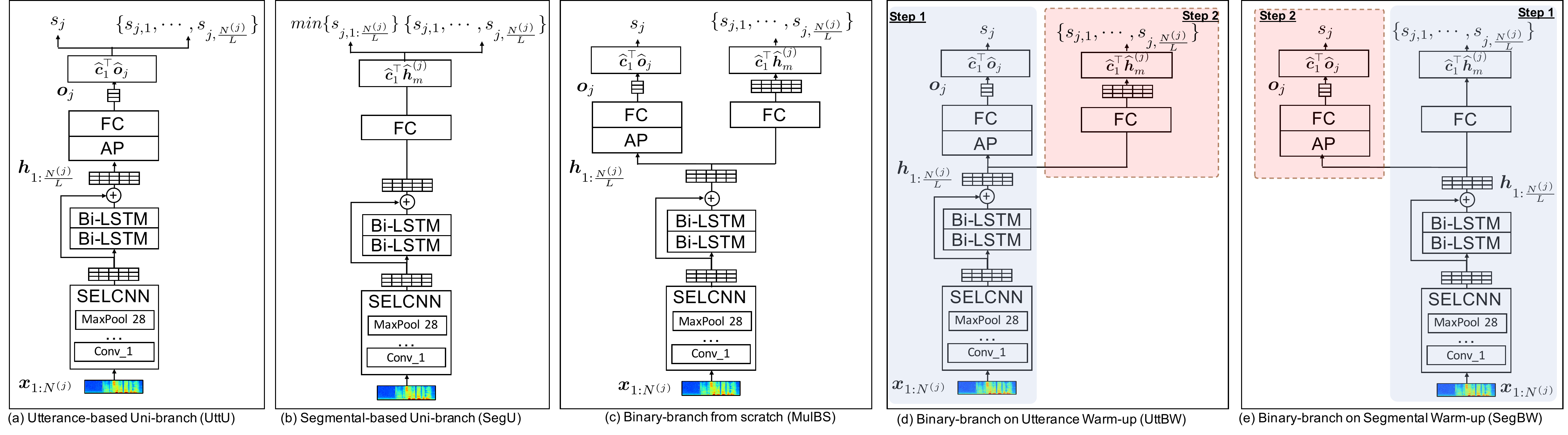}
 \caption{Architecture of multi-task models. Orange dashed boxes indicate components added after pre-training. In output of five models (top of each figure), left part is for utterance level, and right part is for segmental level. }
 \label{fig:multi-model}
 \vspace{-3mm}
\end{figure*}

\textbf{Utterance-basic Model.} Given a feature vector and an utterance-label pair $\{\boldsymbol{x}_{1:N^{(j)}}, y_j\}$, the input $\boldsymbol{x}_{1:N^{(j)}}$ is transformed into $\boldsymbol{h}_{1:\frac{N^{(j)}}{L}}$ by the SELCNN and two Bi-LSTM layers. Utterance loss $\mathcal{L}^{(\text{utt-p2s})}$ is computed during training. Then, we can obtain an utterance-level score $s_j^{(utt)}$ directly during the inference stage. According to the transformation formulas among utterance-level vector $\boldsymbol{o}_j$, bona fide class vector $\boldsymbol{c}_1$ and segmental embedding vectors $\boldsymbol{h}$, the segmental-level score $s_{j,m}^{(utt)}$ can be derived from this utterance-basic model. The utterance-level score for the $j$-th trial $s_j^{(utt)}$ can be decomposed as the sum of segmental scores $s_{j}^{(utt)} = \frac{1}{M_j}\sum_{m=1}^{M_j} s_{j,m}^{(utt)}$, where $M_j=\frac{N^{(j)}}{L}$ is the number of segments produced by strides in the SELCNN. Note that this utterance-basic model is trained using only the utterance-level labels, but it can derive two kinds of scores (the utterance score $s_{j}^{(utt)}$ of the $j$-th trial, and the segment score $s_{j,m}^{(utt)}$ of the $m$-th segment in the $j$-th trial) from the same trained model as described in Table \ref{tab:eq}. 

\textbf{Segmental-Basic Model.}	For the segmental-basic model, we remove the pooling layer from the utterance-basic model to maintain all segmental vectors $\boldsymbol{h}_{1:M_j}$ and use a segmental-level loss function $\mathcal{L}^{(\text{seg-p2s})}$ to conduct segmental training. After that, the segmental-level score $s_{j,m}^{(seg)}$ can be calculated directly. $s_{j}^{(seg)}$ = $\min_{m} {s_{j,m}^{(seg)} }$ \footnote{We also tried using the mean, std, mean - std, and max of segmental score as the utterance score. Min yields the best performance because a segment with a smaller score is more likely to be spoofed, and a spoofed segment declares a spoofed trial.} is treated like an utterance-level score for this segmental-basic model.

In the single-task architecture, models are updated based on one level of labels without any interaction with the other level. It only utilizes limited information from a single level. To adequately utilize both levels of labels, we need the multi-task modeling introduced in the next sub-section.

\subsection{Multi-Task Learning}
In this sub-section, we describe two kinds of multi-task models. Conventional MTL models usually have multiple output branches with one branch per task. Unlike conventional multi-target tasks, the utterance and segmental targets in PartialSpoof are directly related to each other in that a sequence of segmental labels can uniquely determine the utterance label. During inference, the prediction score for the two levels can be derived from each other; that is, they can be predicted from separate branches and from a single model. Thus, we designed the uni-branch and binary-branch MTL architectures for PartialSpoof and updated both architectures using a fused loss function $\mathcal{L} = \mathcal{L}^{(\text{utt-p2s})} + \mathcal{L}^{(\text{seg-p2s})}$.

\subsubsection{Uni-branch Architecture}\label{sec:UttU_SegU}
As discussed, we can transfer the predicted score from one level to another. Based on the relationship between those two levels, we first produce the Uni-branch MTL architecture. Uni-branch MTL architectures are based on single-task models. However, the MTL models produce both the segment- and utterance-level scores, and calculate training losses on both levels simultaneously. We can develop two kinds of uni-branch models leading by utterance (“UttU”) or segment (“SegU”), as shown in Figure \ref{fig:multi-model} (a–b). `U' at the end of the name refers to the uni-branch structure. Mathematical equations can be found in the `Score' column in Table \ref{tab:eq}. For UttU, we specifically use the same neural network processing to extract the segmental score for the segmental vector. For SegU, we use $\min_{m} {s_{j,m} }$ as its utterance score after segmental score prediction. 

\subsubsection{Binary-Branch Architecture}
In the UttU and SegU models in Section \ref{sec:UttU_SegU}, the segmental and utterance-level labels are learned using an entirely-shared neural network model. However, utterance- and segmental-level labels can have conflicting gradients, so the uni-branch structure can be greatly restricted by this conflict. Thus, we attenuate such conflict by applying a binary-branch structure in which the models do not derive scores from each other sequentially but instead simultaneously predict scores from separate branches for each level. As shown in Figure \ref{fig:multi-model} (c), input $\boldsymbol{x}_{1:N^{(j)}}$ is fed into SELCNN and two Bi-LSTM layers to extract an embedding $\boldsymbol{h}_{1:\frac{N^{(j)}}{L}}$. Next, we feed $\boldsymbol{h}_{1:\frac{N^{(j)}}{L}}$ into segmental and utterance branches separately for backpropagation. Finally, we derive the segmental score and utterance score from their respective branches.

\subsection{Pros and Cons}
The uni-branch architecture is more straightforward and simple; it is an upgrade to the basic model and we can generate two-level loss from it without modifying the architecture. But it also aggravates the conflicting gradient problem in MTL. Conflicting gradient is a common problem in MTL \cite{2020YugradientMTL}. In our work, the labels at the segmental and utterance levels for the same signal can provide conflicting information. To mitigate this issue, we build a binary-branch model with independent branches for each type of label and explore it further by applying different training strategies based on training the model from scratch or using warm-up training, which will be introduced in the next section.

\section{Training Strategies for Binary Branch} 
\label{sec:train_multi}
To construct a binary-branch model that enables cooperation between the utterance-level and segmental-level labels, we explore both training a model from scratch and fine-tuning a warm-up model.

\subsection{Training from Scratch}
\label{sec:MulBS}
The simplest way to train a binary-branch model is from scratch. The segmental branch and utterance branch are updated jointly by sharing common layers (SELCNN and two Bi-LSTM) before the pooling layer (as shown in Figure \ref{fig:multi-model} (c)), which we call “\textbf{Mul}ti-task model with \textbf{B}inary-branch training from \textbf{S}cratch” (MulBS).

\subsection{Warm-up} 
While the multi-task model in Section \ref{sec:MulBS} is entirely jointly trained, we consider using one level's pre-trained model to initialize the other level. Thus, we design two warm-up models in this section, as described in Figure \ref{fig:multi-model} (d) and (e). Note that although those warm-up models have a similar architecture to MulBS, their training consists of two stages: (1) the initial warm-up training of a single level and (2) joint fine-tuning of the whole model, including the pre-trained original branch and the expanded branch. Thus, the \textbf{Utt}erance-based \textbf{B}inary-branch \textbf{W}arm-up model (UttBW) and the \textbf{Seg}mental-based \textbf{B}inary-branch \textbf{W}arm-up model (SegBW) can be constructed by initializing from different levels. In UttBW (or SegBW), first, a single utterance (or segmental) model is pre-trained as a warm-up model; next, architecture in another segmental (or utterance) level is added to the position of $\boldsymbol{h}_{1:\frac{N^{(j)}}{L}}$; finally, the initialized branch and the additional branch will be jointly trained in the same manner as in Section \ref{sec:MulBS}. 


\section{Experiments}
\label{sec:exp}
\subsection{Database}
Information about the PartialSpoof\footnote{\scriptsize PartialSpoof Database v1.1: https://zenodo.org/record/5112031\#.YQfkjS2l3iE} \cite{Zhang2021} database used in this work is shown in Table \ref{tab:partialspoof-database-info}.

\begin{table}[th]
\caption{\it Statistic information for bona fide and spoofed samples in PartialSpoof (numbers shown as bona fide/spoof).}
\vspace{2mm}
\label{tab:partialspoof-database-info}
\centerline{
\begin{tabular}{|c|c|c|}
\hline
\textbf{Set}   & \textbf{Utterance Num.} & \textbf{Segmental Num.} \\
\hline \hline
\textbf{Train} & 2,580 / 22,800              & 250,281 / 308,126           \\
\textbf{Dev.}  & 2,548 / 22,296              & 255,261 / 281,690           \\
\textbf{Eval.} & 7,355 / 63,882              & 772,832 / 718,798        \\
\hline
\end{tabular}}
\end{table}

\subsection{Model Description}
We configured the models similarly to \cite{Zhang2021}. We used varied-length sequences of frames of 60-dimensional linear frequency cepstral coefficients (LFCCs) as input $\boldsymbol{x}_{1:N^{(j)}}$, calculated from the 512-point FFT of a 20-ms Hanning window shifted by 10 ms. SELCNN compresses the LFCCs by a factor of 16 along the time dimension as with LCNN. Thus, we computed an embedding vector every 160 ms. We did not use any data augmentation, voice activity detection, or feature normalization.

We used the Adam optimizer for model training. The learning rate was initialized from $3\times 10^{-4}$ and halved every ten epochs. We stopped training when development loss stopped decreasing for 70 epochs. We averaged all the results over six rounds with different random seeds ($10^0 \sim 10^5$). For the warm-up model, we used the best pre-trained model in the development set in the six rounds. We specifically used random seeds equal to $10^6$ and $10^4$ for the pre-trained models of UttBW and SegBW, respectively. All results are reproducible using the same random seeds and environment (PyTorch-1.7 and Nvidia Tesla A100 GPU).

\subsection{Metrics}
For evaluation, we calculated the equal error rate (EER) at both the utterance level and the segmental level by following the ASVspoof 2019 challenge.

\section{Results and Discussions}
\label{sec:res}
\subsection{Baseline Results}
Table \ref{tab:selcnn-seg} compares the results between LCNN-LSTM and SELCNN-LSTM with different reduction-factor configurations in the SE block. We conducted these comparison experiments on a segmental-basic model. SELCNN(2) refers to the results obtained with SELCNN using $r = 2$ (reduction factor in SE block). We can see that SELCNN-LSTM improved compared with LCNN-LSTM, and $r = 2$ performed relatively better. Therefore, SELCNN-LSTM with $r = 2$ serves as a basic model for the rest of this work.

\begin{table}[t]
\caption{\it Comparison of LCNN-LSTM and SELCNN-LSTM with different reductions.}
\vspace{2mm}
\label{tab:selcnn-seg}
\centerline{
\begin{tabular}{|c|cc|cc|}
\hline
\multirow{1}{*}{\textbf{Model } +LSTM} & \multicolumn{2}{c|}{\textbf{Utterance EER (\%)}} & \multicolumn{2}{c|}{\textbf{Segmental EER (\%)}} \\
\textbf{(reduction)}      & \textbf{Dev.}           & \textbf{Eval.}         & \textbf{Dev.}           & \textbf{Eval.}          \\
\hline\hline
\textbf{LCNN}              & 5.01                   & 8.61                  & 6.81                   & 16.21                  \\
\textbf{SELCNN(2)}         & \textbf{4.01}          & \textbf{7.69}         & \textbf{6.38}          & 15.93                  \\
\textbf{SELCNN(4)}         & 5.00                   & 8.12                  & 6.48                   & 16.10                  \\
\textbf{SELCNN(8)}         & 4.49                   & 8.47                  & 6.51                   & 15.96                  \\
\textbf{SELCNN(16)}        & 4.30                   & 8.40                  & 6.54                   & \textbf{15.85}        \\
\hline
\end{tabular}}
\end{table}

\subsection{Comparison of Single-Task and Multi-Task Models}
\label{sec:sig_mul_task}
Single-task and multi-task models are compared in Table \ref{tab:res-pred}. 

\begin{table}[t]
\caption{\it Comparison of single-task and multi-task models.}
\vspace{2mm}
\label{tab:res-pred}
\centerline{
\begin{tabular}{|c|cc|cc|}
\hline
\textbf{Model} & \multicolumn{2}{c|}{\textbf{Utterance EER (\%)}} & \multicolumn{2}{c|}{\textbf{Segmental EER (\%)}} \\
\textbf{types} & \textbf{Dev.}          & \textbf{Eval.}         & \textbf{Dev.}          & \textbf{Eval.}         \\
\hline \hline
\textbf{Utterance} & 3.96                  & 6.33                   & 32.69                  & 44.00                  \\
\textbf{Segment}   & 4.01                  & 7.69                   & \textbf{6.38}                   &\textbf{15.93}                  \\
\hline
\textbf{UttU}           & 8.86                  & 9.96                   & 7.31                   & 20.04                  \\
\textbf{SegU}           & 4.82                  & 7.04                   & 6.82                   & 17.75                  \\
\textbf{MulBS}            & \textbf{2.98}                  & \textbf{5.90}                   & 6.56                   & 17.55                 \\    
\hline
\end{tabular}}
\end{table}

\begin{figure}[thb]
 \centering
 \includegraphics[width=0.95\linewidth]{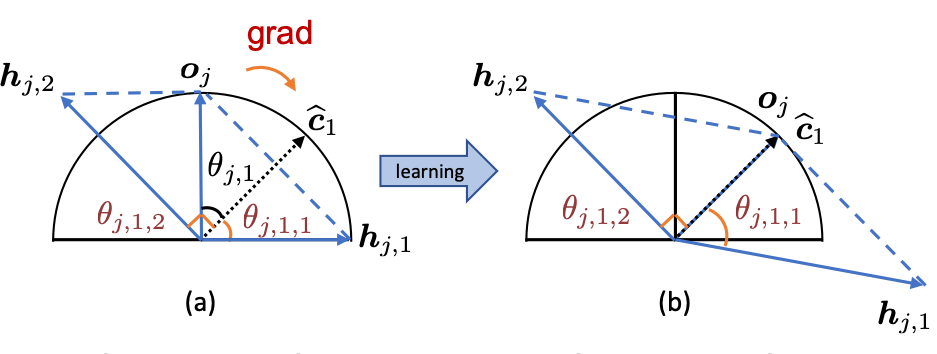}
 \caption{\it Learning process scheme of adapted P2Sgrad loss function $\mathcal{L}^{(\text{utt-p2s})}$ in utterance-basic model that aims to force the utterance-level observation vector $\boldsymbol{o}_j$ closer to the target vector $\hat{c_1}$. Utterance-level observation vector $\boldsymbol{o}_j$ is generated by combining the segmental-level observation vector $\boldsymbol{h}_{j,1}$ and $\boldsymbol{h}_{j,2}$.}
 \label{fig:loss_function}
 \vspace{-3mm}
\end{figure}

It is clear that single-task models (first two rows in Table \ref{tab:res-pred}), especially utterance-basic models, cannot handle detection at both levels. They can perform well at the level for which they were trained, but performance degrades on the other level. Large degradations in the segmental EER appear in the utterance-basic model because a given utterance label cannot uniquely determine the segmental label; therefore, segmental information cannot be utilized during the training stage. Consider the scenario shown in Figure \ref{fig:loss_function}\footnote{This explanation is applicable to models that use length-normalized vectors for classification, e.g., angular softmax and additive-margin softmax.} in which utterance-level embeddings are generated by combining the segmental embedding vectors $\boldsymbol{h}_{j,1}$ and $\boldsymbol{h}_{j,2}$. When we use utterance labels to force the utterance observation vector $\boldsymbol{o}_j$ closer to the correct class vector $\hat{\boldsymbol{c}_1}$ and update the model, the segmental feature vectors $\boldsymbol{h}_{j,1}$ and $\boldsymbol{h}_{j,2}$ are not guaranteed to move towards the target vectors. What matters is the angle between the target class vector and the sum of the feature vectors. The orientation of each segment vector can be changed arbitrarily. In summary, the single-task models have bias because they only consider one level's labels.

\begin{figure}[t]
  \centering
    \includegraphics[width=0.8\linewidth]{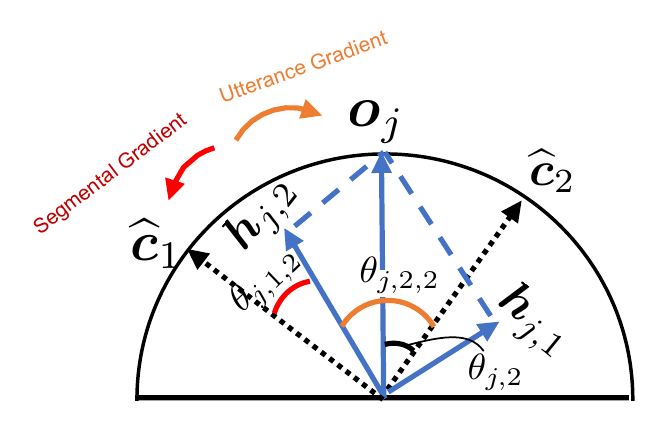}
  \caption{\it Contradiction phenomenon in UttU/SegU}
  \label{fig:contradiction}
  \vspace{-3mm}
\end{figure}

Next, we can look at the results for multi-task models (UttU, SegU, and MulBS). After introducing the segmental labels to the utterance-basic model in UttU, the segmental-level detection performs better but the utterance-level detection degrades—vice versa in SegU compared with the segmental-basic model—indicating that introducing another level's labels is meaningful but sharing the entire neural network can reduce the original level of performance. This is explainable because conflict cases exist, such as how some bona fide feature vectors in a spoofed trial might be updated by two gradients from opposite directions. An example is shown in Figure \ref{fig:contradiction}. Observation vector $\boldsymbol{o}_j$ for the $j$-th trial consists of segmental feature vectors $\boldsymbol{h}_{j,1}$ and $\boldsymbol{h}_{j,2}$. The correct class label for the 2nd segment vector ($\boldsymbol{h}_{j,2}$) is bona fide ($\widehat{\boldsymbol{c}_1}$), but the target vector of global utterance is spoofed ($\widehat{\boldsymbol{c}_2}$), so $\boldsymbol{h}_{j,2}$ is updated by two gradients from opposite directions. Thus, different levels should have their own branches to reduce conflicts between them, like MulBS.

As expected, the binary-branch model MulBS can overcome the weaknesses of both basic single-task models. MulBS performed better on two levels simultaneously than basic models because both utterance-level and segmental-level parameters are considered independently when utterances and segments have their own branches in MulBS; thus, EER is reduced. Because we can treat MulBS as adding an additional segmental branch to the utterance-basic model or as an extension of the segmental-basic model by adding an utterance branch, we can compare results from two points of view: the utterance-basic view and the segmental-basic view. After adding the segmental branch, segmental EER (in dev./eval.) decreased significantly from 32.69\%/44.00\% (utterance-basic) to 6.56\%/17.55\% (MulBS) in the development and evaluation sets. Furthermore, after the adding utterance branch, MulBS improved 23–25\% in utterance EER and equivalent segmental EER over the segmental-basic baseline.

\subsection{Comparison of Training Strategy for Binary-Branch Multi-Task Models}
To further improve the effectiveness of the binary-branch method, we conducted ablation studies on from-scratch and warm-up training strategies for MTL. Results are shown in Table \ref{tab:res-multi}.

The results show that training by fine-tuning a warm-up model obtains a relatively better performance, especially when warming up from the segmental-based model. This indicates that initial training is meaningful and the warm-up step can lead the model closer to the optimal solution, and the fine-grained (segmental) task can help to improve the coarser-grained (utterance) task. Although multi-task models provide improved performance compared to the single-task basic model for utterance-level detection, they did not improve segmental EERs in the evaluation set compared to the simple single-task segmental-basic model.

\begin{table}[t]
\caption{\it Comparison of training strategies for binary-branch models.}
\vspace{2mm}
\label{tab:res-multi}
\begin{tabular}{|c|cc|cc|}
\hline
\textbf{Multi-Task} & \multicolumn{2}{c|}{\textbf{Utterance EER (\%)}} & \multicolumn{2}{c|}{\textbf{Segmental EER (\%)}} \\
\textbf{Types}      & \textbf{Dev.}          & \textbf{Eval.}         & \textbf{Dev.}          & \textbf{Eval.}         \\
\hline \hline

\textbf{MulBS}                      & 2.98                   & 5.90                  & 6.56                   & 17.55                  \\
\textbf{UttBW}                      & 3.30                   & \textbf{5.66}         & 7.10                   & 17.77                  \\
\textbf{SegBW}                      & \textbf{2.53}          & 6.07                  & \textbf{5.78}          &\textbf{16.60}          \\
\hline
\end{tabular}
\end{table}

\section{Conclusions} \label{sec:conclusion}


To investigate the question introduced in Section \ref{sec:intro}: \textit{Can we train one model that can be used for both utterance-level detection and segmental detection?} We first upgrade LCNN to SELCNN by introducing an SE block to LCNN to improve the feature selection ability. Then, we discussed single- and multi-task learning in PartialSpoof based on this new architecture. As expected, the multi-task model was better at detecting spoofing at the utterance and segmental levels simultaneously. Furthermore, MTL helped the model enhance both utterance- and segmental-level detection. Single-task learning is also possible for both levels, but it does not work well. In the multi-task model, the MulBS binary-branch configuration utilized information from the two levels more effectively than the UttU and SegU uni-branch models. 

To further explore the binary-branch model, we investigated three training strategies either from scratch (MulBS) or warm-up (UttBW and SegBW). Detection improved on both utterance and segmental levels after properly considering fine-grained cues (segmental labels), especially in the warm-up model. Those two dependent labels can promote one another through the multi-task binary-branch architecture. Thus, this paper provides a series of multi-task benchmarks for PartialSpoof.


Although our model can detect spoofing at the segmental and utterance levels simultaneously, segmental detection did not show obvious improvement especially in the evaluation set, and the conflict between segmental and utterance labels still exists. Considering all of the above results, segmental detection is still more challenging than utterance-level detection and should be explored further. 

\section{Acknowledgement}

We would like to thank Prof.\ Nicholas Evans and Dr.\ Jose Patino for their comments. This study was supported by the Japanese-French VoicePersonae Project supported by JST CREST (JPMJCR18A6), JST CREST Grants (JPMJCR20D3), MEXT KAKENHI Grants (16H06302, 18H04112, 21H04906), Japan, and the Google AI for Japan program.

\bibliographystyle{IEEEbib}
\bibliography{main.bib}

\end{document}